\documentclass[onecolumn]{svjour3}
\usepackage{graphicx}
\usepackage{url}
\begin{document}

\title{Practical Quantum  Bit Commitment Protocol.}

\author{Ariel Danan and Lev Vaidman}
\institute{ Raymond and Beverly Sackler School of Physics and Astronomy
 Tel-Aviv University, Tel-Aviv 69978, Israel \\ Email: vaidman@post.tau.ac.il}

\maketitle

\begin{abstract}
A quantum protocol for bit commitment the security of which is based
on technological limitations on nondemolition measurements and
long-term quantum memory is presented.
\end{abstract}

\vspace{.6cm}


Quantum cryptography, which began with Wiesner's idea \cite{wiesner1983conjugate} almost forty years ago, has reached the stage of commercial key distribution devices \cite{magiq,id,tosh}.
Apart from celebrated protocols for quantum key distribution
\cite{bennett1984quantum,ekert1991quantum}, there has been hope for an unconditionally secure
quantum bit commitment protocol \cite{brassard1993quantum}. In such a protocol, Alice
decides about the value of a bit and gives it encrypted to Bob.
Later, Alice gives Bob the key which enables him to read the bit. In
an ideal bit commitment scheme, Bob, without the key, cannot get any information about
the bit and Alice unable to change it. However, it has been shown that
unconditionally secure quantum bit commitment is
impossible due to the so called ``EPR attack'', in which Alice gives
Bob one particle of the Einstein-Podolsky-Rosen pair instead of
giving him a particle in a pure quantum state \cite{lo1997quantum,mayers1997unconditionally}. There is
now a widespread consensus that the proof is indeed correct and we do
not contest it.

In this paper we propose a quantum bit commitment protocol which,
although not unconditionally secure, is very secure in practice. Its
security  relies on the current technological limitations on the
storage of qubits and on nondemolition measurements of qubits
carried by photons. We believe that for any foreseen future there will be a large gap between fidelity of immediate demolition measurements and fidelity of measurements of photon qubits performed after a nondemolition measurement of the presences of the photon and its long storage. This will ensure the security of the protocol.
Current classical bit commitment methods which rely on the
difficulty in performing certain computational tasks will most
probably not survive very long due to developments in computational
power. But more importantly, it is possible for the classical
methods to be broken even today by a clever mathematician
discovering an effective search algorithm.

The proposed bit commitment protocol is based on the secure quantum
key distribution protocol. In fact, various quantum key distribution
methods can be modified to fit the protocol, but we will discuss just
the most famous and the simplest protocol proposed by Bennett and
Brassard at 1984 (BB84) \cite{bennett1984quantum}. The main advantages of the
proposed method are its simplicity and the possibility of its
immediate realization using the technology of today.
In the protocol, Alice makes a commitment, but it is Bob who sends a
number of qubits (photons) to Alice.\\

The protocol consists of the following steps:
\begin{enumerate}
  \item  Bob sends a number (which specifies the level of security) of photons to Alice. The photons are prepared randomly in one of the four polarization states of the BB84 protocol (L,R,X,Y) and sent at random times. Bob keeps the record of when and what he has sent to Alice. (For practical reasons, Bob might send faint pulses instead of single photons. He records then the polarization of each pulse.)
  \item Alice measures all the photons in one out of two bases and this choice manifests her commitment. She announces immediately the times of detection of the photons.
  \item At the opening stage Alice reveals the bit commitment, i.e., the basis of each measurement and its outcome. The outcomes have to match the polarization states of the qubits sent by Bob in the same basis.
\end{enumerate}

Since Alice reveals the outcomes of her measurements only at the
opening stage, Bob gains no information about Alice's commitment.
The times of arrival of the photons yield no information, provided
Alice arranges her detection devices carefully, avoiding dependence
of the probability of detection on the basis of photon polarization
measurement.

An effective cheating strategy for Alice, which is allowed by
physical laws, consists of a nondemolition measurement of the time
of arrival of each photon followed by storage of the qubit
information encoded in the photon's polarization in some stable
physical system. Then, just before the opening stage, Alice can make
her choice and measure the qubits in the appropriate basis. Today,
however, in spite of recent efforts \cite{munro2005high}, both nondemolition
measurement and long-term storage of qubits are out of technological
reach.

The system which will allow running this bit commitment protocol is
one which provides BB84 secure secret key distribution. A source
based on heralded single photons will be very good, but even a weak
pulse source will serve the purpose. Bob should send weak enough
pulses of known polarization, such that the rate of detection of
pairs of photons in a pulse, provided there are no losses in the
channel, is much smaller than the actual rate of photon detection by
Alice.

 A well designed device which allows secure key distribution
will be at the same time secure against a similar attempt of Alice
to postpone the commitment.
A measurement in one basis destroys the eigenstates of the other basis, so Alice cannot measure polarization
in both bases. In order to cheat, Alice needs to do something equivalent to what Eve does in the key distribution protocol: She should try to announce only the detection of photons that arrive in
pairs. If the rate of detecting pairs $p_2$ is too law, the best strategy of Alice is to measure polarization of single photons  randomly in one of the intermediate Breidbart's basis \cite{bennett1992experimental}.
\begin{eqnarray}
 |V_{1}\rangle=\sin\frac{\pi}{8}|X\rangle+i\cos\frac{\pi}{8}|Y\rangle, \ \
 |U_{1}\rangle=\cos\frac{\pi}{8}|X\rangle-i\sin\frac{\pi}{8}|Y\rangle,\\
 |V_{2}\rangle=\sin\frac{\pi}{8}|X\rangle-i\cos\frac{\pi}{8}|Y\rangle, \ \
 |U_{2}\rangle=\cos\frac{\pi}{8}|X\rangle+i\sin\frac{\pi}{8}|Y\rangle.
\end{eqnarray}
This will lead to  the  quantum bit error rate (QBER) of Alice's measurement relative to Bob's chosen basis of:
\begin{equation}\label{3}
    QBER=\sin^{2}\frac{\pi}{8}\cdot100\%\cong15\%,
\end{equation}
and to the expected 50\% for the orthogonal basis (due to random choice of Alice's bases). Thus, the total QBER when Alice cheats using ideal devices is:
\begin{equation}\label{4}
   QBER\cong(1-p_2)\cdot15\%.
\end{equation}
A practically secure bit commitment should have lower error rate.

The current proposal has much in common with  the (non-secure)  bit commitment method described by Bennett and Brassard \cite{bennett1984quantum}. In their method the setup is essentially the same, but the protocol is different. Bob, instead of sending random polarizations (L,R,X,Y), records and sends random polarization either among the pair (L,R) or the pair (X,Y).
{\it His} choice manifests the commitment he makes to Alice. Now, Alice measures the photons in random bases and records the results. At the opening stage, Bob reveals the commitment and provides the polarizations of the photons he sent, which should coincide with polarizations measured by Alice in the same basis. The cheating strategy for Bob consists of sending one photon from an EPR pair and keeping the second photon until he really wants to make the commitment. Since today long-term qubit memory is unavailable, this method is also practically secure. However, its security is significantly weaker than the proposed method. The difference is that in this method Bob can perform an unlimited number of trials for creating an EPR pair in which one particle is a photon and the other is a stable qubit. (``Stable qubit'' might be, e.g., a quantum state of a two-level atom.) Therefore, in order to cheat, it is sufficient to have a technology that is at all capable of creating such pairs, however inefficient. In contrast, in the proposed scheme we have to be able to measure
the time of arrival of each photon as well as transfer its qubit to a stable system with finite efficiency. The required efficiency depends on the losses of the actual quantum channel and Alice's ability to reduce these losses. Also, the allowed error rate in Alice's procedure is specified by the error rate in the transmitted
qubits, which is low in current experiments.

We have learned recently that Damgard et al. proposed to switch  the roles of Alice and Bob already at 2005 \cite{damgard2005cryptography}. However, what is missing in  their proposal is a random timing (or faint pulses) which eliminates requirement of technology of nondemolition measurements. The security of their method relies solely on the absence of reliable storage of photon qubits.
\pagebreak

In the proposed method, a qubit may be encoded in various ways in the quantum state of a photon, not necessarily in its polarization. A promising possibility for bit commitment, when there is a large
distance between Alice and Bob, is using time bin encoding \cite{PhysRevLett.62.2205,PhysRevLett.82.2594}. The difficulty with long distance protocols is that most of them, for solving the problem of disturbances, use two-line two-way quantum channels. This allows to cancel uncontrollable disturbances introduced by the two lines. The two-way protocol cannot be easily adapted for the proposed bit commitment, so the time bin encoding which can be implemented in a single quantum line is advantageous.

It is important to note that, even for short distances, quantum bit commitment is of great importance since it achieves a task which cannot be achieved using classical means. For the purpose of secret key distribution, Alice and Bob, when they are not far from each other, can just meet and bring one another the ``one time pad''. But secret bit commitment is something Alice and Bob cannot perform even if they are sitting together. Unconditionally secure classical bit commitment requires a trusted party. Thus, we return to the situation in which the difference between classical and quantum cryptographic methods boils down to the former being based on the security offered by computational complexity, and the latter on the technological feasibility of quantum nondemolition measurements and long-term quantum memory.

Note that the error rate in quantum communication for short distances is very low using current technology, therefore the proposed bit commitment method is highly secure. The gap between the error rate of an immediate demolition measurement and that of a postponed measurement (i.e., following a nondemolition measurement
of the photon's time of arrival and qubit storage) will remain significant for a long time, providing high security for bit commitment between proximate parties.

To test the proposed protocol we have built a toy quantum communication system, using fiber optic components. The qubit encoding was based on the proposal of Townsend et al. \cite{townsend1993enhanced} for quantum cryptography systems. The latter is based on  splitting a photon into two time bins, using unbalanced Mach Zehnder interferometer (UMZI), and modulating the relative phase between them for the qubit encoding. The four qubit states are then:
\begin{eqnarray}\label{1}
|X\rangle=\frac{1}{\sqrt{2}}(|A\rangle+|B\rangle), \nonumber \\
|Y\rangle=\frac{1}{\sqrt{2}}(|A\rangle-|B\rangle),\\
|L\rangle=\frac{1}{\sqrt{2}}(|A\rangle+i|B\rangle),\nonumber \\
|R\rangle=\frac{1}{\sqrt{2}}(|A\rangle-i|B\rangle),\nonumber
\end{eqnarray}
where $|A\rangle$ and $|B\rangle$ are the outputs of the short arm and the long arm of the UMZI, respectively. These time bins are then transmitted to Alice's identical UMZI, and the outcome of her measurement is the result of interference of the \emph{long-short + short-long} time bins. Alice can control her measurement basis by modulating the relative phase in her UMZI.
\pagebreak

\begin{figure}
\begin{center}
  \includegraphics [scale=0.37]{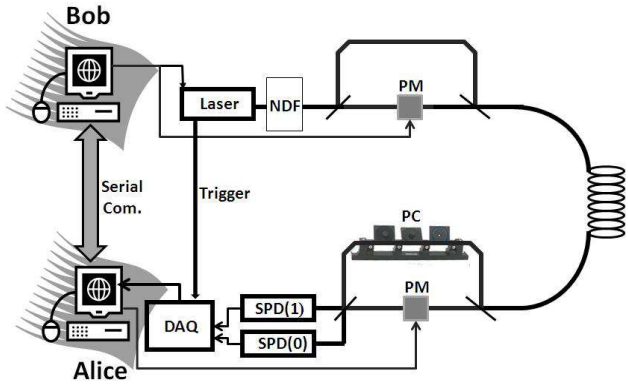}
\end{center}
  \caption{A schematic diagram of the QBC system. DAQ-data acquisition card; NDF-natural density filter; PM-piezoelectric mount; SPD-single photon detectors; PC-polarization controller. }\label{SETUP}
\end{figure}

We use a nanosecond pulse laser (Toptica iPulse) as a photon source, and a single mode optical fiber as a quantum channel (Fig. \ref{SETUP}). The laser pulse is attenuated to the single photon regime (0.2 photons per pulse on average) using neutral density filters.  Subsequently, the faint pulse is coupled into the fiber-optic UMZI. The UMZI itself is constructed of 2X2 fiber couplers, each with a 50/50 splitting ratio. The long arm of the UMZI consists of a 15m-long single mode fiber, and the short arm is a free-space delay line, adjustable by means of a high-resolution piezoelectric actuator. The later is controlled by Bob's computer, and allows encoding of 4 different phase shifts, associated with the four different qubits. The encoded double faint pulses are transmitted through a single mode fiber to Alice's receiver.

The encoded double pulses enter Alice's identical fiber-optic UMZI. The free space delay line is again controlled by a piezoelectric actuator, which allows the setting of the measurement basis by Alice's computer.
A polarization controller in Alice's UMZI allows compensation for polarization rotations in the fibers, and ensures that the interfering pulses (\emph{long-short + short-long}) reach the final fiber coupler with the same polarization. The exit ports of Bob's UMZI are coupled to two single-photon detectors (SPD-idquantique id100), whose outputs represent the two possible outcomes of the measurement (1 and 0). The SPD signals are sampled by a data acquisition card, which is trigged by a synchronization signal sent by Bob. This allows selective detection of the photons which arrive during the time window corresponding to the interference pulse.\pagebreak

\begin{figure}
\centering
  \includegraphics [scale=0.3]{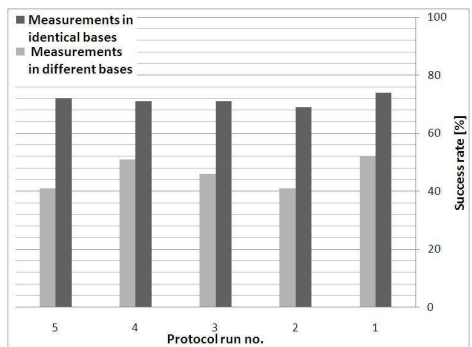}
  \caption{Opening stage results of five QBC protocol runs with 200 sent qubits each. The black (gray) columns represents measurement success rate when the sent qubit was (was not) in the commitment basis.}\label{results}
\end{figure}

Figure \ref{results} shows typical results obtained with the toy system described above. The data clearly shows a consistently higher success rate for measurements with identical bases compared to measurements with different bases. However, the experimental QBER is higher (i.e. fidelity is lower) than that required for compliance with the theoretical security limit (15 percent QBER). The reason for this is a low interference visibility, due to limitations of the laser switching. Nevertheless, these results show that the protocol is feasible with current technology quantum communications. We believe that a more robust implementation would be able to compete with classical methods based on computational complexity.\\

We thank Paul Kwiat, Tal Mor, and  Stephen Wiesner for useful discussions and we thank Yoav Linzon and Shimshon Bar-Ad for help with the experiment.
This work has been supported in part by the Binational Science Foundation Grant No. 32/08, the Israel M.O.D Grant No.03214921 and the Wolfson Family Charitable Trust.

\bibliography{qbc_bib}

\end {document}